\def\cm{cm$^{-1}$}
\documentclass{epl}
\title{Charge Localization due to RKKY Interaction\\ in the Spin Glass  AuFe}
\shorttitle{Charge Localization in Spin Glasses}

\author{B. Gorshunov\inst{1,2},
\thanks{email: groshunov@ran.gpi.ru}
A.S. Prokhorov\inst{2}, S. Kaiser\inst{1}, D. Faltermeier\inst{1}, S.
Yasin\inst{1},  M. Dumm\inst{1},  N. Drichko\inst{1,3},  E.S.
Zhukova\inst{2}, I.E. Spektor\inst{2}, S. Vongtragool\inst{4,5}, M.B.S.
Hesselberth\inst{4},  J. Aarts\inst{4}, G.J. Nieuwenhuys\inst{4,5},
and M. Dressel\inst{1} } \shortauthor{B. Gorshunov \etal} \institute{
  \inst{1} 1.~Physikalisches Institut, Universit{\"a}t
Stuttgart, 70550 Stuttgart, Germany\\
  \inst{2} Prokhorov General Physics Institute, Russian Academy of Sciences, Moscow, Russia\\
  \inst{3} Ioffe Physico-Technical Institute, St.\ Petersburg, Russia\\
  \inst{4} Kamerlingh Onnes Laboratory, Leiden University, Leiden, The Netherlands\\
  \inst{5} Res. Department Condensed Matter, Paul Scherrer Institute, Villigen, Switzerland}

\begin{document}

\maketitle

\begin{abstract}
Measurements of electrodynamic response of of spin glass AuFe films in
comparison with pure gold films are performed at frequencies from
0.3~THz (10 \cm) up to 1000 THz (33\,000 \cm) using different
spectroscopic methods. At room temperatures the spectra of pure gold
and of AuFe are typically metallic with the scattering rate of carriers
in AuFe being significantly enlarged due to scattering on localized
magnetic moments of Fe ions. In the spin-glass phase of AuFe at $T=5$~K
a pseudogap in the conductivity spectrum is detected with the magnitude
close to the Ruderman-Kittel-Kasuya-Yosida (RKKY) energy for AuFe:
$\Delta_{\rm RKKY}\approx 2.2$~meV. The origin of the pseudogap is
associated with partial localization of electrons which mediate the
RKKY interaction between localized magnetic Fe centers.
\end{abstract}

\section{Introduction}
Phenomena in spin glasses represent one of the central topics of modern
solid state physics; they are of fundamental interest and also have a
variety of possible applications \cite{Mydosh93}. The spin-glass state
is realized in intermetallic alloys, for instance, when ions of a
magnetic metal (like Fe, Mn) are introduced in small amounts into the
matrix of non-magnetic noble metals (like Au, Ag, Cu, Pt). The local
magnetic moments interact co-operatively with each other via the
conduction electrons by the agency of the indirect
Ruderman-Kittel-Kasuya-Yosida (RKKY) exchange interaction
\cite{Freeman72}. Magnitude and sign of the interaction depend on the
distance between impurities. Combined with the spatial disorder this
provides conditions for a spin-glass state.

Among the exceptional properties of spin glasses compared to other
magnetic materials is the temperature behavior of their magnetic
susceptibility, which reveals a kink at a certain temperature $T_f$
(the freezing temperature) whose shape and position depend on the
magnitude and alternation frequency of the probing field
\cite{Binder86}. Spin glasses possess magnetic memory: the magnitude of
magnetization created by an external magnetic field below $T_f$ depends
on the pre-history of the system. Typical for a spin-glass state are
relaxational phenomena with characteristic times which at low
temperatures can by far exceed the duration of the experiment. In spite
of the large number of theoretical and experimental investigations,
there is still no generally accepted consensus on the nature of the
spin-glass state and the majority of properties of spin glasses remain
not fully understood \cite{Mydosh93,Binder86,Fischer91,Mezard87}.

Since the RKKY interaction plays a fundamental role in the physics of
spin glasses, the behavior of the subsystem of free electrons should be
intimately linked to the formation and stabilization of the spin-glass
phase. The magnitude of the RKKY interaction depends on the electronic
mean free path, as was first shown by de Gennes \cite{deGennes62}.
Thus, investigating the characteristics of conduction electrons gives
insight into  the peculiar physics of spin glasses. The most direct way
to study the properties of delocalized electrons is provided by
electrical transport experiments. Immediately following the first works
on spin glasses, the electrical resistance of ``classical'' systems
like AuFe, CuMn, AuMn, and AuCr has been investigated in a detailed and
systematic way as a function of temperature, magnetic field and
concentration of magnetic centers
\cite{Loram70,Ford70,Mydosh74,Ford76,Campbell82}. It was shown that the
magnetic contribution to the electrical resistivity $\rho(T)$ reveals a
$T^{3/2}$ temperature dependence at the lowest temperatures and a $T^2$
dependence close to $T_f$; at elevated temperatures $T>T_f$ there is a
broad maximum in $\rho(T)$ which is due to  a competition between Kondo
and RKKY interactions in the subsystems of electrons and magnetic
moments. Existing theories encounter serious difficulties to reproduce
the temperature behavior of the resistivity in broad intervals of
temperatures and impurity concentrations \cite{Binder86}. Certain
difficulties are also caused by deviations from Matthiessen's rule at
elevated temperatures.

Fundamental information on the properties of the electronic subsystem
can be obtained by optical spectroscopy, which for instance allows one
to extract such characteristics of free carriers as mechanisms of
scattering and relaxation, energy gaps and pseudogaps in the density of
states, localization and hopping parameters, size and granularity
effects in thin conducting films \cite{DresselGruner02}. However, to
our knowledge, there are no data published on optical spectroscopy of
spin glasses. The reason may be purely technical: since these materials
are highly conducting, almost like regular metals, it is practically
impossible to measure their electrodynamic properties by standard
spectroscopical techniques, especially in the far-infrared range and at
even lower frequencies where effects of interactions of mobile
electrons with localized spins and between these spins should reveal
themselves. Here we present the first measurements of the
electrodynamic response of the spin-glass compound AuFe in a broad
range of frequencies with an emphasis on the THz range corresponding to
energies of the radiation quanta which are close to the RKKY binding
energy.

\section{Experimental Techniques}

For the measurements we have chosen the well-studied spin-glass
compound AuFe. A set of films with different thicknesses and Fe
concentrations was prepared. The high purity metals were co-sputtered
onto a high-resistive Si substrate (size $10\times 10$~mm$^2$,
thickness about 0.5~mm). Before the argon sputter gas was admitted the
equipment was pumped down to UHV conditions ($10^{-7}$~torr) to prevent
oxidation of the films during fabrication. The films were analyzed
using Rutherford backscattering and electron microprobe analysis; the
thickness and composition was homogeneous. In this paper we concentrate
on the results obtained for an AuFe film with 6 at.\%\ of Fe and about
50 nm thickness. We also measured a pure Au film of the same thickness
prepared under the same conditions.

For the THz investigations a coherent source spectrometer
\cite{Kozlov98} was used which operates in the frequency range from 30
GHz up to 1.5 THz (1 - 50~\cm). This range is covered by a set of
backward-wave oscillators as powerful sources of radiation whose
frequency can be continuously tuned within certain limits. In a
quasioptical arrangement the complex (amplitude and phase) transmission
and reflection coefficients can be measured at temperatures from 2~K to
1000~K and in a magnetic field up to 8~Tesla if required. Dynamical
conductivity of Au and AuFe films was directly determined from THz
transmissivity and reflectivity spectra in a way we have used for
measurements  other conducting films, like heavy fermions
\cite{Dressel02} or superconductors \cite{Pronin96}.

In order to complete our overall picture, the samples were optically
characterized up to the ultraviolet. The room temperature experiments
were conducted on the same Au and AuFe films as the THz investigations.
In the infrared spectral range ($600- 7000$~\cm), optical reflectivity
R$(\omega)$ measurements were performed using an infrared microscope
connected to a Bruker IFS 66v Fourier transform spectrometer. An
aluminum mirror served as reference, whose reflectivity was corrected
by the literature data \cite{Palik85}. A Woollam vertical variable
angle spectroscopic ellipsometer (VASE) equipped with a Berek
compensator was utilized to measure in the energy range between
5000~\cm\ and 33\,000~\cm\ with a resolution of 200~\cm\ under multiple
angles of incident between 65$^{\circ}$ and 85$^{\circ}$. From the
ellipsometric measurements we obtain the real and imaginary parts of
the refractive index which then allow us to directly evaluate any
optical parameter like the reflectivity $R(\omega)$ or the conductivity
$\sigma(\omega)$.

\section{Experimental Results and Discussion}
\begin{figure}
\centering\resizebox{1\columnwidth}{!}{\includegraphics{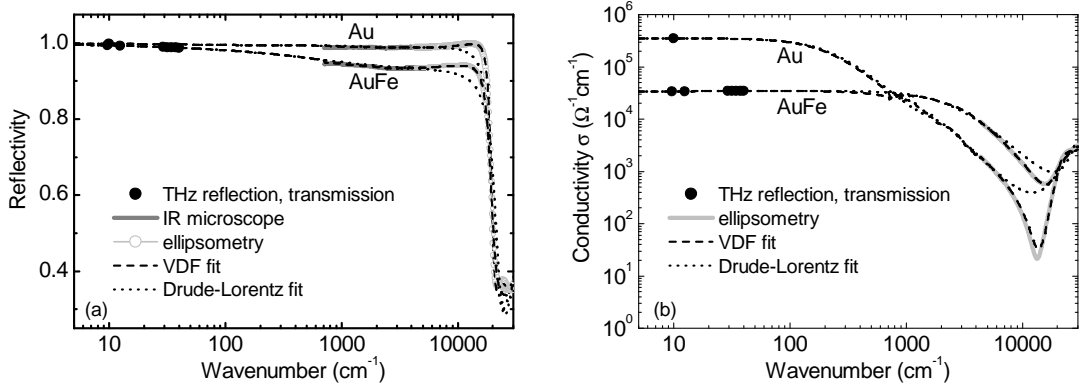}}
 \caption{(a) Room temperature reflectivity and (b) conductivity of Au and AuFe (6 at.\%\ Fe) films.
 The full circles below 50~\cm\ are from THz transmission and reflection measurements.
 The solid curves correspond to the IR reflection measurements.
 The reflectivity calculated from the ellipsometric measurements is given be grey open circles and lines.
 The thick lines in the conductivity spectra between 5000~\cm\ and 33\,000~\cm\ are directly
 calculated from ellisometric data.
 The dotted lines are the fits by a simple Drude-Lorentz model.
 The dashed lines indicate a combined fit of reflectivity and conductivity by an advanced
 Drude-Lorentz model based on the variational dielectric function
 introduced in \protect\cite{Kuzmenko05}.}\label{fig:opticalspectra}
\end{figure}
To analyze the frequency dependent transport, we first consider the
room temperature results displayed in Fig.~\ref{fig:opticalspectra}. It
is seen that the spectra for both Au and AuFe are metallic
\cite{DresselGruner02}: the reflectivity reveals a characteristic
plasma edge around 20000~\cm and the conductivity $\sigma(\omega)$ is
only weekly frequency dependent at low frequencies and quickly drops
between $10^3$~\cm\ and $10^4$~\cm. The increase of the conductivity at
even higher frequencies (above $2\cdot 10^4$~\cm) is caused by
electronic interband transitions. In order to extract the microscopic
characteristics of charge carriers, we fitted the spectra by the Drude
model of conductivity \cite{DresselGruner02}:
$\hat{\sigma}(\omega)=\sigma_{\rm dc}/(1-i\omega\tau)$, where
$\sigma_{\rm dc}$ denotes the dc conductivity and  $\tau=1/(2\pi
c\gamma)$ the relaxation time and $\gamma$ the relaxation rate of
charge carriers (c is the speed of light). The higher frequency
interband transitions were roughly modelled by additional Lorentz
oscillators. The results of the fit are presented by dashed lines in
Fig.~\ref{fig:opticalspectra}, the parameters are summarized in
Table~\ref{tab:1}. As indicated by the dotted lines, both the
reflectivity and conductivity spectra can be perfectly reproduced by a
more advanced procedure based on the variational analysis of the
optical reflectivity and conductivity spectra introduced by Kuzmenko
\cite{Kuzmenko05}. The low-frequency conductivity is smaller and the
scattering rate of carriers larger (more than ten times) in AuFe
compared to Au. Obviously, the differences should be ascribed to
additional magnetic scattering of electrons in AuFe. Although the
plasma frequency is not affected when Fe is diluted in Au, the
distribution of spectral weight (as measured by the center of gravity,
for instance) is shifted to higher energies.
\begin{table}
\caption{Drude parameters of the free charge carriers in Au and AuFe (6
at.\%\ Fe) films obtained by a Drude-Lorentz fit to the room
temperature spectra of Fig.~\ref{fig:opticalspectra} with the dc
($\omega\rightarrow 0$) conductivity $\sigma_{\rm dc}$, the scattering
rate $\gamma$, scattering time $\tau$,  and the plasma frequency
$\omega_p=\sqrt{4\pi ne^2/m}$ with $n$ and $e$ being the concentration
and the charge of the carriers, and $m$ their effective mass.}
\label{tab:1}
\begin{tabular}{ccccc}
\hline\noalign{\smallskip}
Film & $\sigma_{\rm dc}$ ($\rm\Omega^{-1}$cm$^{-1}$)& $\gamma$ (\cm)  & $\tau$ (s)& $\omega_p$ (\cm) \\
\noalign{\smallskip}\hline\noalign{\smallskip}
Au &  350\,500&~~236 & $2.25\times 10^{-14}$ & 70\,450 \\
AuFe  & ~~33\,600 & 2445 & $2.17\times 10^{-15}$ & 70\,100 \\
\noalign{\smallskip}\hline
\end{tabular}
\end{table}

\begin{figure}
\centering\resizebox{0.5\columnwidth}{!}{\includegraphics{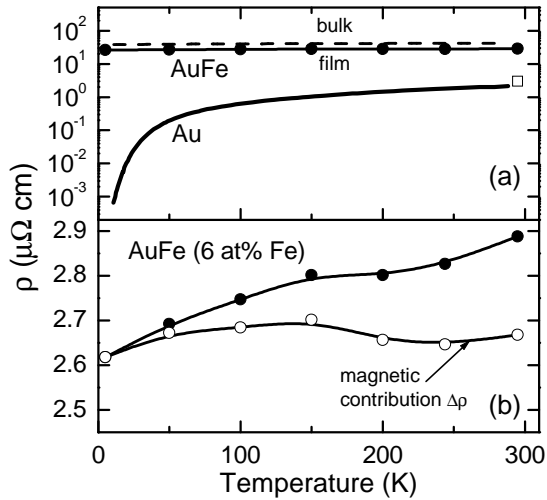}}
 \caption{(a) Temperature dependent ac and dc resistivities of Au and AuFe.
 The solid dots correspond to ac resistivity $\rho(\omega)$ = 1/$\sigma(\omega)$ of AuFe film (6 at.\%\ Fe)
 measured at 35~\cm\ (1.05~THz).
 The dashed line refers to the dc resistivity obtained for bulk AuFe with 5 at.\%\ of Fe \cite{Mydosh74}.
 The solid line indicates the dc resistivity of bulk gold \protect\cite{Mydosh74}.
 The square shows the ac resistivity of Au film at 35 \cm (1.05 THz).
(b) The temperature dependence of the magnetic contribution to the ac
resistivity evaluated by $\Delta\rho(T)=\rho_{\rm AuFe}(T)-\rho_{\rm
Au}(T)$ is shown by the open circles; for comparison the ac resistivity
of the AuFe film is re-plotted on a linear scale (solid
dots).}\label{fig:temperaturedependence}
\end{figure}
The temperature dependence of the transport characteristics is
presented in Fig.~\ref{fig:temperaturedependence}. The upper panel
compares the ac resistivity $\rho(\omega)$ = 1/$\sigma(\omega)$ of the
AuFe (6 at.\%\ Fe) and Au films to the dc resistivity of a bulk AuFe
(with slightly different Fe concentration of 5 at.\%) and of pure bulk
Au samples (data from Ref.~\cite{Mydosh74}).

First, it is obvious that at all temperatures the resistivity of our
AuFe film is very close to that of the bulk material: for example, at
room temperature $\rho_{\rm AuFe}(\rm film) \approx 30~\mu\Omega$cm and
$\rho_{\rm AuFe}(\rm bulk)\approx 40~\mu\Omega$cm. The same holds for
the pure Au samples: $\rho_{\rm Au}(\rm film) \approx 3~\mu\Omega$cm
and $\rho_{\rm Au}(\rm bulk)\approx 2~\mu\Omega$cm. This agreement
indicates the very good quality of our thin films and that there are
basically no effects on their ac electrical properties connected with a
possible granular structure. The same conclusion is also drawn from the
measurements of the freezing temperatures $T_f\approx 25$~K of our AuFe
film which appears to be basically the same as those for bulk samples.

Furthermore, it is seen from Fig.~\ref{fig:temperaturedependence} that
at all temperatures the resistivity of AuFe is much larger than the
resistivity of Au; the difference increases when cooling down. This is
a consequence of scattering of the charge carriers on magnetic
impurities that prevails over phonon scattering in the entire
temperature range. The resistivity $\rho(T)$ reveals a broad feature
around $100-150$~K which is ascribed to the interplay of Kondo and RKKY
regimes \cite{Ford70,Mydosh74,Ford76,Campbell82}. At high temperatures
thermal excitations exceed the RKKY energy of interacting impurities
and a Kondo-like scattering of electrons on independent magnetic
moments dominates. This leads to a weak increase of the magnetic
contribution to the resistivity $\rho_{\rm mag}$ upon cooling as
demonstrated by the open circles in
Fig.~\ref{fig:temperaturedependence}b where the difference $\Delta\rho
=\rho_{\rm  AuFe} -\rho_{\rm  Au}=\rho_{\rm mag}$ is plotted. At low
temperatures the RKKY interaction between magnetic moments starts to
surmount and causes a noticeable suppression of the magnetic
contribution to the resistivity. Assuming Matthiessen's rule
\cite{Bass72}, the scattering rate of electrons due to magnetic
interaction in AuFe at $T=300$~K can be calculated as $\gamma_{\rm mag}
= \gamma_{\rm AuFe} - \gamma_{\rm Au} \approx 2210$~\cm.

\begin{figure}[b]
\centering\resizebox{0.5\columnwidth}{!}{\includegraphics{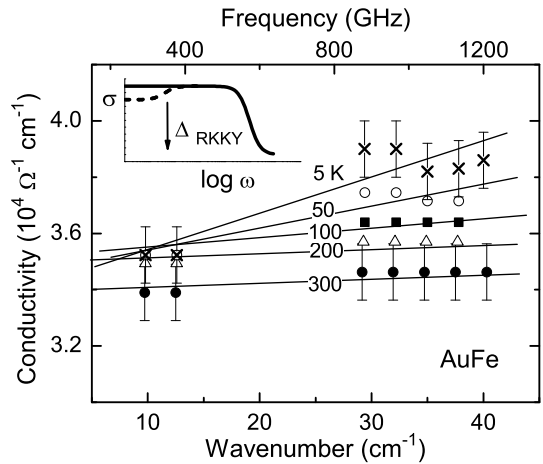}}
 \caption{Terahertz conductivity spectra of the AuFe film (6 at.\%\ Fe)
 at  various temperatures. For $T<100$~K a positive dispersion indicates localization effects.
 The lines are guides to the eye. The inset shows schematically how the Drude-like conductivity spectrum of free electrons (solid line) is modified by a gap-like structure (dashed line) due to
the RKKY interaction between Fe magnetic moments mediated by these electrons.}\label{fig:THzspectra}
\end{figure}
In the low temperature regime, also the frequency dependent
conductivity of AuFe is distinct from the room temperature behavior as
demonstrated in Fig.~\ref{fig:THzspectra} where the THz spectra of
$\sigma(\omega)$ for AuFe are presented. Above approximately 100~K,
$\sigma(\omega)$ is basically frequency independent in accordance with
the Drude model which predicts a constant conductivity for frequencies
much smaller than the scattering rate \cite{DresselGruner02}, as
sketched in the inset of Fig.~\ref{fig:THzspectra} by the solid line;
the scattering rate for AuFe equals 2445~\cm\ at 300~K (Table
\ref{tab:1}), i.e., far above the range of frequencies presented in
Fig.~\ref{fig:THzspectra}. This changes drastically for $T<100$~K, when
the conductivity $\sigma(\omega)$ increases towards high frequencies.
We associate this conductivity dispersion with a pseudogap which
appears in the free electron excitations when the RKKY interaction
between Fe centers mediated by the conduction electrons sets in. In a
simple picture the electrons which participate in the RKKY interaction
can be regarded as being to some extent bound to (or localized between)
the corresponding pairs of magnetic moments, as long as the thermal
energy $k_BT$ does not exceed this ``binding energy'' which should be
of order of the RKKY interaction $\Delta_{\rm RKKY}$. This will lead to
a corresponding reduction of the dc conductivity and also of the ac
conductivity for frequencies below $\Delta_{\rm RKKY}/\hbar$. At higher
frequencies, $\omega>\Delta_{\rm RKKY}/\hbar$, the electrons will no
longer be affected (and localized) by the RKKY interaction and hence
the conductivity $\sigma(\omega)$ should increase around $\Delta_{\rm
RKKY}/\hbar$ to approach the unperturbed value. In other words, one
would expect a gap-like feature to appear in the conductivity spectrum,
as depicted by the dashed line in the inset of
Fig.~\ref{fig:THzspectra}. The RKKY energy  is approximately given by
the freezing temperature $T_f$ \cite{Schilling76}, which for AuFe (6
at.\%\ of Fe) is about 25 K \cite{Mydosh74,Canella72}, yielding
$\Delta_{\rm RKKY}\approx k_BT_f\approx 2.2$~meV.  For the
characteristic frequency we then obtain $\Delta_{\rm
RKKY}/\hbar\approx17$~\cm\ (510 GHz). This falls just in the range
where the dispersion of the conductivity of AuFe in the spin-glass
state is observed. The effect amounts to approximately 10\%, meaning
that about one tenth  of the conduction electrons participate in the
RKKY interaction.

According to our picture of spin-glass systems, the conduction
electrons experience two effects from the RKKY interaction mediated by
these electrons. On one hand, a decrease of the resistivity is commonly
observed while cooling below $T_f$ because the RKKY correlations
between magnetic moments progressively suppress the Kondo-type
scattering. On the other hand, a certain fraction of carriers is
increasingly bound to the magnetic moments by participating in the RKKY
interaction and is thus taken out of the conduction channel. The
competing character of the two effects is clearly seen in
Fig.~\ref{fig:THzspectra}: while cooling down, the gap-like feature
appears on top of a background conductivity which increases basically
at all shown frequencies. In order to verify our assumptions, a
comprehensive study is required on various spin-glass materials with
different freezing temperature and consequently different $\Delta_{\rm
RKKY}$.

\section{Conclusions}
The optical spectra of pure gold and spin glass AuFe (6 at.\%\ Fe)
films have been investigated in a broad frequency range from 10~\cm\ up
to 33\,000~\cm\ using three different spectroscopic techniques. At
ambient temperature the microscopic charge-carrier parameters in pure
gold and in AuFe are determined. For the spin glass AuFe the scattering
rate of the carriers is significantly enlarged due to their interaction
with localized magnetic moments. At reduced temperatures ($T<100$~K)
when the RKKY interaction gains importance as the spin-glass state is
formed, a pseudogap feature in the optical conductivity spectrum is
detected of a magnitude close to the RKKY energy in AuFe. We associate
the origin of the pseudogap with partial localization of those
electrons which are involved in the RKKY interaction between magnetic
moments.
\section{Aknowledgements}  The work was supported by the Russian foundation
for Basic Research, grant N06-02-16010-a and the Deutsche
For\-schungs\-ge\-mein\-schaft (DFG). We also thank the Foundation for
Fundamental Research of Matter (FOM).

\end{document}